\documentclass[preprint,onecolumn,showkeys,amsmath,amssymb,prl]{revtex4}
\usepackage{epsf}
\usepackage{dcolumn}
\usepackage{bm}
\everymath{\displaystyle}
\begin{document}

\title{Comparative analysis of direct and "step-by-step"
Foldy–Wouthuysen transformation methods}
\author{Alexander J. Silenko}
\affiliation{Research Institute for Nuclear Problems, Belarusian State University, Minsk 220030, Belarus}

\date{\today}

\begin {abstract}
Relativistic methods for the Foldy–Wouthuysen transformation of the "step-by-step" type already at the
first step give an expression for the Hamilton operator not coinciding with the exact result determined by
the Eriksen method. The methods agree for the zeroth and first orders in the Planck constant terms but
do not agree for the second and higher-order terms. We analyze the benefits and drawbacks of various
methods and establish their applicability boundaries.
\end{abstract}


\maketitle

\section{Introduction}

Because the Foldy–Wouthuysen (FW) representation \cite{FW} has some unique properties, it holds a special
place in quantum mechanics. In this representation, quantum mechanical operators for relativistic particles
in an external field have the same form as in the nonrelativistic quantum theory. In particular, the position
operator \cite{NW} and momentum operator are equal to $\bm r$ and $\bm p=-i\hbar\nabla$, and the polarization operator for
spin-1/2 particles is expressed by the Dirac matrix $\bm \Pi$. In other representations, much more cumbersome
formulas are used to write these operators (see
\cite{FW,JMP}). The relations between the operators in the FW
representation are analogous to the relations between the corresponding classical quantities. The simple
form of operators corresponding to classical observables is a great advantage of this representation. These
properties of the FWrepresentation allow using it successfully for passing to the semiclassical approximation
and the classical limit of relativistic quantum mechanics \cite{FW,CMcK}. We note that the Hamiltonian and all other
operators are diagonal in two spinors (block-diagonal) in this representation.

When the FW representation is used, the passage to the classical limit is usually accomplished by
simply replacing the operators in the expressions for the Hamiltonian and in the operator equations for
the dynamics with the corresponding classical quantities. The possibility of such a replacement, explicitly
or implicitly used in practically all works devoted to the relativistic FW transformation, was recently
rigorously proved in \cite{JINRLett12}. This possibility radically simplifies interpreting the basic quantum mechanics
equations, especially in the relativistic case.

For practical purposes, the Hamilton operator in the FW representation must be derived up to terms
of the order $\hbar^2$. The contribution to the Hamilton operator provided by the scalar electric and magnetic
polarizability and for spin
$s>1/2$ particles also by the quadrupole interaction and tensor electric and
magnetic polarizability has exactly this order of magnitude. We note that the terms characterizing the
polarizability are of the second order in the field. As an example showing the importance of such interaction
analysis for modern experimental physics, we mention the need to take the tensor polarizability into
account in experiments seeking the electric dipole moment of the deuteron. In these experiments, the tensor
polarizability of the deuteron can be successfully measured (see
\cite{BarJPhys2008,PRC2009,TenPolJul} and the references therein).

The importance of the FW representation for modern quantum mechanics and elementary particle
physics makes the problem of passing to this representation quite relevant. The basic methods for the
passage are direct, which allows passing to the FW representation in a single transformation, and "step by
step" (iterative methods). Here, we analyze these methods comparatively and establish their applicability
boundaries.

We use the $c=1$ system of units. At the same time, the Planck constant $\hbar$ is included in the equations.

\section{Methods for passing to the FW representation}

Passing to the FW representation is a highly nontrivial problem. It was noted relatively early that
such a passage is by no means identical to bringing Hamiltonian to a block-diagonal form (see \cite{JMPcond}
and
the references therein). In particular, as shown in
\cite{EK}, even the classical method, developed by Foldy
and Wouthuysen \cite{FW}, strictly speaking, does not lead to this representation. In \cite{FW}, passing to the blockdiagonal
form is done "step by step," in successive iterations, each of which results in removal of odd
(non-block-diagonal) terms of the highest order. But the operator $ U_{FW}$ of the \emph{exact} FW transformation $ U_{FW}
~(\Psi_{FW}=U_{FW}\Psi)$ for spin-1/2 particles should satisfy the Eriksen condition \cite{E}
\begin{eqnarray}
\beta U_{FW}=U^\dag_{FW}\beta, \label{pdnvEri}
\end{eqnarray}
where $\beta$ is a Dirac matrix. With the operator $U_{FW}$ represented in the exponential form
\begin{eqnarray}
U_{FW}=\exp{(iS)} \label{expfEri}
\end{eqnarray}
condition (\ref{pdnvEri}) is equivalent to the requirement that the exponent $S$ should be Hermitian and odd \cite{EK}. In
view of the Hausdorff theorem \cite{Hausdorff}
\begin{eqnarray}
\exp(A)\exp(B)=\exp{\left(A+B+\frac12[A,B]+{\rm higher~ order~
commutators}\right)},\nonumber\\ \exp(A)\exp(B)\neq
\exp(B)\exp(A).\label{EriKorl}
\end{eqnarray}
If $A=iS_1$ and $B=iS_2$, where
$S_1$ and $S_2$ are odd Hermitian operators, then the
operator $[A,B]$
is odd, and $\exp(A)\exp(B)$ does not satisfy Eriksen condition
(\ref{pdnvEri}) \cite{EK}. Therefore, the classical FW
method \cite{FW} and other methods of "step-by-step" type do not satisfy the oddness condition for the operator $S$
and can consequently provide only an approximate passage to the FW representation. Because Dirac
matrices do not commute, the operator $[S_1,S_2]$ can be of the same order as $S_2$. In this case, even the second
iteration is useless. But formula (\ref{EriKorl}) does not allow estimating the error coming from iterative methods
quantitatively.

Eriksen found a general form for the operator of the exact transformation to the FW representation
(the FW transformation) in the static case
\cite{E}. The initial Hamiltonian for spin-1/2 particles can be
represented in the general form
\begin{equation} {\cal H}_D=\beta m+{\cal E}+{\cal
O},~~~\beta{\cal E}={\cal E}\beta, ~~~\beta{\cal O}=-{\cal
O}\beta, \label{2eq3} \end{equation} where ${\cal E}$ and ${\cal O}$
are even and odd operators. The operator found by Eriksen is defined by the expression
\begin{equation}
U_{FW}=\frac12(1+\beta\lambda)\left[1+\frac14(\beta\lambda+\lambda\beta-2)\right]^{-1/2},
\label{eq3EI}
\end{equation}
where $\lambda={\cal H}_D\left/\sqrt{{\cal H}_D^2}\right.$. The quantity
$\lambda$ takes the respective values $+1$ and $-1$ for states with positive
and negative energies. It is important that
\cite{E}
\begin{equation}\lambda^2=1, ~~~ [\beta\lambda,\lambda\beta]=0, \label{eq3.12}
\end{equation}
and the operator $\beta\lambda+\lambda\beta$ is even:
\begin{equation}[\beta,(\beta\lambda+\lambda\beta)]=0.\label{eqEII}
\end{equation}
Even operators are block-diagonal and do not mix upper and lower spinors.
Formula (\ref{eq3EI}) can also be written in the form \cite{JMPcond}
\begin{equation}
U_{FW}=\frac{1+\beta\lambda}{\sqrt{(1+\beta\lambda)^\dag(1+\beta\lambda)}}.
\label{eq3.15}
\end{equation}
The two operator factors in the radicand commute.

The operator $U_{FW}$ annihilates the respective lower or upper spinor of any eigenfunction of Dirac
Hamiltonian for positive or negative energy. This transformation is done in one step.

Other direct methods for the FW transformation were developed in
\cite{St,neznamov,neznamovEChAYa}. We limit ourself to
considering the Eriksen method because it was thoroughly justified in \cite{VJ}.

But it easily seen that it is problematic to effectively use the Eriksen method with the goal of obtaining
relativistic formulas for particles in an external field because general formula (\ref{eq3EI}) is extremely cumbersome
and contains square roots of Dirac matrices. The most general expression for the Hamiltonian operator
in the FW representation, found by the Eriksen method in \cite{VJ}, represents this operator as a series of
relativistic corrections in powers of the operators ${\cal E}/m$ and ${\cal
O}/m$.
This expansion gives a good solution
of the problem for nonrelativistic particle velocities. In particular, it can be used for electrons in atoms
(except heavy atoms) because
$|v/c|\sim\sqrt Z\alpha\ll1$. But the Eriksen method, for example, does not allow
passing to the FW representation for fast particles moving in external fields (in accelerators and storage
rings). For ${\cal O}^2/m^2\approx\bm p^2/m^2\ge1$, the relativistic correction series does not converge at all. We therefore
cannot use the Eriksen method to find compact relativistic expressions for the Hamilton operator in this
representation.

In contrast to the Eriksen method, some of the iterative methods give the sought relativistic expressions
\cite{JMP,B,PRA,Gos,Gos2,GosHM,GosM}. We note that the method developed in
\cite{PRA} is applicable to particles with
any spin. Because all these methods are approximate, we must determine their applicability boundaries.
Obviously, the simplest, most reliable way to do this is to compare relativistic Hamiltonians in the FW representation
obtained by methods of the "step-by-step" type with the exact power series given in \cite{VJ}. This
problem is extremely important, in particular, because the terms proportional to the second derivatives of
the field potentials and to the squared field strengths are checked. As indicated above, taking them into
account can be necessary in considering effects due to scalar and tensor polarizability.

Here, we compare results obtained by the three methods developed in \cite{B}, in \cite{JMP,PRA} and in
\cite{Gos,Gos2,GosHM,GosM}. These methods have the most fundamental justification among the methods of the "step-by-step" type. We
then compare these results with results of the Eriksen method and draw conclusions about the precision of
the FW transformations given by these methods.

\section{Comparison of results obtained by different methods of the
"step-by-step" type}

To compare the results, certainly, we need to use Eqs. (\ref{2eq3}) as the initial and formal expression for the
Hamiltonian in the FW representation in terms of the operators ${\cal E}$ and ${\cal
O}$. The transformed Hamiltonian
was represented in precisely this form in \cite{EK,E,VJ,DeVries}. Although some fairly general problems were
considered in
\cite{B,Gos,Gos2,GosHM,GosM}, the above form was not used there. On the other hand, a concrete form of the
Hamiltonian operator in the FW representation was obtained in
\cite{JMP} in the weak-field approximation, i.e.,
with only first-order terms in field potentials and their derivatives taken into account. We therefore first
determine this operator with the precision needed for comparison, using general equation (31) in \cite{JMP}. This
equation has the form
\begin{equation}
{\cal H}_{FW}=\beta\epsilon+ {\cal E}'+\frac
14\beta\left\{\frac{1} {\epsilon},{{\cal O}'}^2\right\},~~~
\epsilon=\sqrt{m^2+{\cal O}^2}, \label{eq14} \end{equation} where
$\left\{\dots,\dots\right\}$
denotes the anticommutator and ${\cal
E}'$ and ${\cal O}'$ are the even and odd operators after the first step
of the transformation, defined by the expressions \cite{JMP}:
\begin{equation}  \begin{array}{c}
{\cal E}'={\cal E}-\frac14\left[\frac{\epsilon+m}
{\sqrt{2\epsilon(\epsilon+m)}},\left[\frac{\epsilon+m}
{\sqrt{2\epsilon(\epsilon+m)}},{\cal F}\right]\right] \\
+\frac14\left[\frac{\beta{\cal O}}
{\sqrt{2\epsilon(\epsilon+m)}},\left[\frac{\beta{\cal O}}
{\sqrt{2\epsilon(\epsilon+m)}},{\cal F}\right]\right], \\
{\cal O}'=\frac{\beta{\cal O}}{\sqrt{2\epsilon(\epsilon+m)}} {\cal
F}\frac{\epsilon+m}{\sqrt{2\epsilon(\epsilon+m)}} \\ -
\frac{\epsilon+m}{\sqrt{2\epsilon(\epsilon+m)}}{\cal
F}\frac{\beta{\cal O}}{\sqrt{2\epsilon(\epsilon+m)}},
\end{array} \label{eq28} \end{equation} where
$${\cal
F}={\cal E}-i\hbar\frac{\partial}{\partial t}.$$

Deriving the Hamiltonian in the FW representation with a fixed precision in the Planck constant and
writing the initial Hamiltonian formally as in (\ref{2eq3}), we must use the following a priori information. Every
commutation of the operator $\bm p$ with some function $f(\bm r)$ of coordinates (e.g., with the scalar potential),
compared with the product $\bm pf(\bm r)$,
adds a factor of the order $\hbar/S_0$, where $S_0$ is some value with the
dimension of action. When the condition $\lambda_B\ll l$ of small de Broglie wavelength $\lambda_B=\hbar/p$ compared with
the characteristic size $l$ of the external field inhomogeneity region (or the particle localization region) is
satisfied, the commutator of the operators is smaller in order of magnitude than their product:
\begin{equation}
\frac{|[\bm p,f(\bm r)]|}{|pf(\bm
r)|}\sim\frac{\hbar}{lp}=\frac{\lambda_B}{l}\ll1.
\label{eqdeBro}\end{equation}

To find the order of $S_0$, we use an estimate of the means of the corresponding Hamiltonian terms. It
follows from
(\ref{eqdeBro}) that $S_0=lp$. For particle beams in accelerators and storage rings, this value is equal to the
angular momentum ($S_0=rp=L$, where $r$ is the ring radius), and the condition
$\hbar/S_0\ll1$ is automatically
satisfied. Certainly, this condition is not satisfied in all cases. In particular, very often (whenis automatically
satisfied. Certainly, this condition is not satisfied in all cases. In particular, very often (when $rp\sim\hbar$, where $r$ is the electron orbit radius), it does not hold for electrons in atoms. At the same time, the compact
character of an interaction (electro-weak, for example), as the analysis in \cite{TMP97} shows, does not preclude a
correct description of relativistic effects using the FW transformation.

One more standard reason why the operators ${\cal
O}$ and ${\cal E}$ do not commute is the noncommutativity of
different components of the kinetic momentum operator $\bm{\pi}=\bm p-e\bm A:~
[\pi_i,\pi_j]=iee_{ijk}B_k$. In this case, we
have the estimate for relativistic particles
$$\frac{|[\pi_i,\pi_j]|}{|\pi_i\pi_j|}\sim\frac{\hbar}{S_0}=\frac{|e|\hbar
B}{\epsilon^2},$$ where $\epsilon$ is the total kinetic energy including the rest energy. This absence of commutativity usually leads
to spin-dependent terms appearing in the Hamilton operator in the FW representation, and $S_0=\epsilon^2/(|e|B)$ is significantly greater than $\hbar$ as a rule.

In the general case, the order of magnitude of the commutator $[{\cal O},{\cal E}]$
is determined by the commutator
of the operator $\bm p$ 
with a function of coordinates or by their commutators with matrices contained in the
operator ${\cal O}$ if the operator ${\cal E}$ contains even matrices
$\bm\Sigma$ or $\bm\Pi$. For example, for spin-1/2 particles in
homogenous electric and magnetic fields \cite{JMP}, we have
$$
{\cal E}=e\Phi-\mu'\bm {\Pi}\cdot\bm{B}, ~~~{\cal
O}=\bm{\alpha}\cdot\bm{\pi}+ i\mu'\bm{\gamma}\cdot\bm{E}. $$ In this case,
\begin{eqnarray}[{\cal O},{\cal
E}]=ie\hbar\bm{\alpha}\cdot\bm{E}-2\beta\gamma^5\mu'\bm{\pi}\cdot\bm{B}
-2i\gamma^5{\mu'}^2\bm{E}\cdot\bm{B}.
\label{eq22tmp}\end{eqnarray} Because $\mu'=(g-2)e\hbar/(4mc)$,
the second term in the right-hand side of (\ref{eq22tmp}),
resulting from the noncommutativity
of the matrices $\bm\alpha$ and $\bm{\Pi}$,
is also proportional to $\hbar$. If the order of magnitude of ${\cal E}$
is determined by
the scalar potential and $|g-2|\sim1$, then $S_0=m|\Phi|/B$ for this term. With $B\sim E$ for relativistic particles,
it has the same order of magnitude as the first term.

The operator ${\cal O}$, being odd (non-block-diagonal), contains the Dirac matrices $\gamma_1,\gamma_2$, and $\gamma_3$. In
accordance with the properties of these matrices, multiple commutators of the form $[{\cal O},[{\cal O},\dots[{\cal O},{\cal E}]\dots]]$ have the order of magnitude $\hbar/S_0$ with respect to the operator
product ${\cal O}{\cal O}\dots{\cal O}{\cal E}$ with the factor $\hbar$ already appearing in the result of the first commutation. In contrast, commutators of the forms $[{\cal O}^2,[{\cal
O},{\cal E}]]$, $[{\cal O}^2,[{\cal O}^2,{\cal E}]]$, and $[[{\cal
O},{\cal E}],{\cal E}]$ are of the order $(\hbar/S_0)^2$ (with respect to the product of the operators appearing
in them). This property follows because ${\cal
O}^2$ and ${\cal E}$ are even operators and do not contain the Dirac matrices
$\gamma_1,\gamma_2$, and $\gamma_3$ which are non-block-diagonal.

Determining the order of magnitude, we indicate the smallest possible degree in $\hbar$. For example, the
commutator $[{\cal O},{\cal E}]$ with the nominal order $\hbar/S_0$ can, as in (\ref{eq22tmp}), contain terms of the orders
$(\hbar/S_0)^2, (\hbar/S_0)^3, \dots$, can have an order higher than one in $\hbar/S_0$ (see Sec. \ref{example}), or can just be zero.

In \cite{JMP}, the expressions for ${\cal H}_{FW}$
were computed only for concrete problems, and the weak-field approximation
was used. But using general formulas
(\ref{eq14}),(\ref{eq28}), we can easily determine the form of this operator
up to terms of the order $(\hbar/S_0)^2$:
\begin{equation}\begin{array}{c}
{\cal H}_{FW}=\beta\epsilon+ {\cal E}-\frac 18\left\{\frac{1}
{\epsilon(\epsilon+m)},[{\cal O},[{\cal O},{\cal
F}]]\right\}\\+\frac {1}{64}\left\{\frac{2\epsilon^2+2\epsilon
m+m^2} {\epsilon^4(\epsilon+m)^2},[{\cal O}^2,[{\cal O}^2,{\cal
F}]]\right\}\\-\frac {1}{16}\beta\left\{\frac{1}
{\epsilon^3},\left([{\cal O},{\cal F}]\right)^2\right\}+\frac
{1}{64}\beta\left\{\frac{1} {\epsilon^5},\left([{\cal O}^2,{\cal
F}]\right)^2\right\}. \end{array} \label{eq14gen}
\end{equation}

To compare results obtained by different methods of the "step-by-step" type, we consider Dirac particles
in an inhomogeneous electrostatic field. In this case, in formula (\ref{eq14gen}),
obtained by the method developed in \cite{JMP,PRA}, we
have ${\cal E}=e\Phi$ and ${\cal O}=\bm\alpha\cdot\bm p$, where $\Phi$
is the scalar potential and $\bm\alpha$ is a Dirac matrix. We note that the
expression for ${\cal E}$ and
${\cal O}$ is nontrivial, which excludes the possibility of random coincidence. Equation (\ref{eq14gen}) becomes
\begin{equation}\begin{array}{c}
{\cal H}_{FW}=\beta\epsilon'+ e\Phi\\
+\frac{e\hbar}{8}\left\{\frac{1}
{\epsilon'(\epsilon'+m)},\bigl[\bm\Sigma\!\cdot\!(\bm
p\!\times\!\bm E)-\bm\Sigma\!\cdot\!(\bm E\!\times\!\bm
p)+\hbar\Delta\Phi\bigr]\right\}\\-\frac
{e\hbar^2}{16}\left\{\frac{2{\epsilon'}^2+2\epsilon' m+m^2}
{{\epsilon'}^4(\epsilon'+m)^2},(\bm p\cdot \nabla)(\bm p\cdot
\nabla)\Phi\right\}\\+\frac {e^2\hbar^2}{16}\beta\left\{\frac{1}
{{\epsilon'}^3},\bm E^2\right\}-\frac
{e^2\hbar^2}{64}\beta\left\{\frac{1} {{\epsilon'}^5},\left(\bm
p\cdot\bm E+\bm E\cdot\bm p\right)^2\right\},
\end{array} \label{eq14ele}
\end{equation}
where $\epsilon'=\sqrt{m^2+\bm p^2}$ and $\bm E=-\nabla\Phi$. 

The result shown above coincides with corresponding expressions obtained in \cite{B,GosM} (the magnetic
field was also taken into account in the first of them). An analysis of the methods being compared shows
that this coincidence is perfectly natural. All three methods are relativistic, and the FW transformation is
realized in them with the same scheme. The transformation operators are chosen such that they annihilate
the total odd operator
$({\cal O},{\cal O}',{\cal O}'',\dots)$ under the condition that it commutes with the total even operator $({\cal E},{\cal E}',{\cal
E}'',\dots)$, in the first and subsequent stages ("steps") and also with the operator $\epsilon$, the operators $\epsilon''$, etc. in the second and subsequent stages. Such commutativity is absent in general, and we therefore obtain a converging series of corrections defining the transformed Hamilton operator. All three methods should lead
to equivalent results because their difference consists in using different quantum mechanics formalisms. In
the method proposed in \cite{B}, unitary transformations are not used. It is based on the Moyal quantum mechanics
(see \cite{OsMol} and the references therein), where quantum mechanical operators are associated with "classical"
distributions in the phase space. Quantum mechanical evolution is described using Moyal brackets, corresponding
to commutators in ordinary quantum mechanics, while observable quantities are characterized
by functions on the phase space. In the original method proposed in
\cite{Gos,Gos2,GosHM,GosM}, unitary transformations are
also not used. The quantum mechanical system Hamiltonian is represented as a matrix $H_0(\bm P,\bm R)$, whose
elements are operators depending on the pair of canonical variables
$\bm P$ and $\bm R$. This method determines
a diagonalization procedure based on formal series in powers of the Planck constant $\hbar$ and can be used
for a wide class of Hamiltonians, for which Berry phase corrections are essential. Unlike these methods,
the ordinary form of quantum mechanical operators was used in \cite{JMP,PRA}
and the FW transformation is
accomplished using unitary transformations.

But the agreement of the results obtained by the three methods only means that in the framework of
these methods, the initial Hamiltonian is transformed to the same representation but not that this representation
is indeed the FW representation. We also note that this agreement is an additional confirmation
of the correctness of the quantum mechanics formalisms used in \cite{B,Gos,Gos2,GosHM,GosM}.

Some other methods for the relativistic FW transformation, in particular, the method proposed
in \cite{TMP,TMP97}, based on one variant of the exclusion method developed by Akhiezer, Berestetskii, and Landau \cite{BL,AB}, also lead to correct results in computing terms of the zeroth and first orders in $\hbar/S_0$. The
method used in \cite{BliokhK}. also gives correct expressions. At the same time, some relativistic methods, including
those proposed for passing to the FW representation, yield incorrect results even in the first order in $\hbar/S_0$ (see examples in \cite{JMPcond}).

Comparing the three methods under consideration leads to the following conclusion. The most cumbersome
computations are needed in the method in \cite{Gos,Gos2,GosHM,GosM}, while using the traditional mathematical
apparatus of quantum mechanics (in \cite{JMP,PRA}) allows passing to the FW representation most easily. Writing
the initial Hamiltonian in form (\ref{2eq3}) and using general formula (\ref{eq14gen})
simplifies the computations even more.
We note that the easiest way to derive the relativistic Hamilton operator in the traditional operator framework
is to compute with formula
(\ref{eqfin}) below, where, compared with (\ref{eq14gen}), terms of the order
$(\hbar/S_0)^2$, not
computable by iterative methods, are omitted. But we should note that the need for substantially more
computation when using the methods developed in \cite{B,Gos,Gos2,GosHM,GosM} to a large extent results from the need to
switch between different quantum mechanics formalisms in the initial and final expressions. We also note
importance of the results obtained in \cite{Gos,Gos2,GosHM,GosM,Gosgrap} (also see the references therein) in studying effects due
to Berry phases.

\section{Comparison of Hamiltonians obtained by the Eriksen method
and methods of the "step-by-step" type}

To determine the precision of methods of the "step-by-step" type, we must compare Eq. (\ref{eq14gen}) with
the Hamilton operator expansion in the FW representation in powers of the operators ${\cal E}$ and ${\cal O}$ and their
products containing sufficiently large powers obtained by the Eriksen method. Such an expansion was
found in \cite{VJ}. But the result there, obtained using symbolic computer computations, is represented in a
form very inconvenient for this comparison. Reducing the Hamiltonian operator found in \cite{VJ} by writing it
via multiple commutators leads to the expression
\begin{equation}\begin{array}{c}
{\cal H}^{(1)}_{FW}=\beta\left(m+ \frac {{\cal
O}^2}{2m}-\frac{{\cal O}^4}{8m^3}+\frac{{\cal
O}^6}{16m^5}-\frac{5{\cal O}^8}{128m^7}\right)\\+{\cal
E}-\frac{1}{128m^6}\left\{(8m^4-6m^2{\cal O}^2+5{\cal O}^4),[{\cal
O},[{\cal O},{\cal
E}]]\right\}\\+\frac{1}{512m^6}\left\{(2m^2-{\cal O}^2),[{\cal
O}^2,[{\cal O}^2,{\cal E}]]\right\}\\
+\frac{1}{16m^3}\beta\left\{{\cal O},\left[[{\cal O},{\cal
E}],{\cal E}\right]\right\}-\frac{1}{32m^4}\left[{\cal
O},\left[[[{\cal O},{\cal E}],{\cal E}],{\cal
E}\right]\right]\\+\frac{11}{1024m^6}\left[{\cal O}^2,\left[{\cal
O}^2,[{\cal O},[{\cal O},{\cal E}]]\right]\right]+A_{24},
\end{array} \label{eq12erk}
\end{equation}
where
$$\begin{array}{c} A_{24}=\frac{1}{256m^5}\beta\Biggl(24\left\{{\cal O}^2, ([{\cal O},{\cal E}])^2\right\}-
11\left ([{\cal O}^2,{\cal E}]\right )^2\\-14\left\{{\cal O}^2,
\left [[{\cal O}^2,{\cal E}],{\cal E}\right]
\right\}-4\left[{\cal O},\left[{\cal O},\left[[{\cal O}^2,{\cal
E}],{\cal E}\right] \right] \right]\\+\frac92 \left[\left[{\cal
O},\left[{\cal O},\left[{\cal O}^2,{\cal E}\right]\right] \right],{\cal
E}\right] +\frac52 \left [{\cal O}^2,\left[{\cal
O},\left[[{\cal O},{\cal E}],{\cal E}\right] \right]
\right]\Biggr).
\end{array} $$

In $A_{24}$, the first and second subscripts indicate the respective numbers of ${\cal E}$ and ${\cal O}$ operators in the product.
The two preceding terms in Eq. (\ref{eq12erk}), defining $A_{32}$ and part of
$A_{16}$, have the order of magnitude $(\hbar/S_0)^3$ and can be discarded. Neglecting these operators and also terms of the third degree in $\hbar$ in $A_{24}$, we obtain
\begin{equation}\begin{array}{c}
{\cal H}^{(1)}_{FW}=\beta\left(m+ \frac {{\cal
O}^2}{2m}-\frac{{\cal O}^4}{8m^3}+\frac{{\cal
O}^6}{16m^5}-\frac{5{\cal O}^8}{128m^7}\right)\\+{\cal
E}-\frac{1}{128m^6}\left\{(8m^4-6m^2{\cal O}^2+5{\cal O}^4),[{\cal
O},[{\cal O},{\cal
E}]]\right\}\\+\frac{1}{512m^6}\left\{(2m^2-{\cal O}^2),[{\cal
O}^2,[{\cal O}^2,{\cal E}]]\right\}\\
+\frac{1}{16m^3}\beta\left\{{\cal O},\left[[{\cal O},{\cal
E}],{\cal E}\right]\right\}+
\frac{1}{256m^5}\beta\Biggl(24\left\{{\cal O}^2, ([{\cal O},{\cal
E}])^2\right\}\\- 11\left ([{\cal O}^2,{\cal E}]\right
)^2-14\left\{{\cal O}^2, \left [[{\cal O}^2,{\cal E}],{\cal
E}\right] \right\}\Biggr).
\end{array} \label{eq14erk}
\end{equation}

To compare (\ref{eq14gen})
with the exact solution, we must write it in the static case and also represent it as a
series of relativistic corrections in powers of the operators ${\cal E}/m$ and ${\cal
O}/m$. In this case, it becomes
\begin{equation}\begin{array}{c}
{\cal H}_{FW}={\cal H}^{(2)}_{FW}=\beta\left(m+ \frac {{\cal
O}^2}{2m}-\frac{{\cal O}^4}{8m^3}\right.\\ \left.+\frac{{\cal
O}^6}{16m^5}-\frac{5{\cal O}^8}{128m^7}\right)+{\cal
E}\\-\frac{1}{128m^6}\left\{(8m^4-6m^2{\cal O}^2+5{\cal
O}^4),[{\cal O},[{\cal O},{\cal
E}]]\right\}\\+\frac{1}{512m^6}\left\{(10m^2-19{\cal O}^2),[{\cal
O}^2,[{\cal O}^2,{\cal E}]]\right\}\\
-\frac{1}{8m^3}\beta\left([{\cal O},{\cal
E}]\right)^2+\frac{1}{32m^5}\beta\left([{\cal O}^2,{\cal
E}]\right)^2.
\end{array} \label{eqgen}
\end{equation}

If we consider the nonstationary case, change the criterion for estimating the magnitude of the terms
in Eqs. (\ref{eq14gen}) and (\ref{eqgen}), and represent these equations as an expansion in powers of ${\cal
F}/m,$ and ${\cal O}/m$, limiting
ourself to terms of the third order in the inverse mass, then the indicated relations lead to an expression
coinciding with those obtained in \cite{EK} by the classical FW method \cite{FW}:
\begin{equation}\begin{array}{c}
{\cal H}_{FW}={\cal H}^{(3)}_{FW}=\beta\left(m+ \frac {{\cal
O}^2}{2m}-\frac{{\cal O}^4}{8m^3}\right)+{\cal
E}\\-\frac{1}{8m^2}[{\cal O},[{\cal O},{\cal F}]]
-\frac{1}{8m^3}\beta\left([{\cal O},{\cal F}]\right)^2.
\end{array} \label{eqoFW}
\end{equation}
This also shows that the results obtained by the different iterative methods agree. But we note that for
terms of higher order in the inverse mass in the power series in ${\cal F}/m$ and ${\cal O}/m$ , the results derived using
the classical method \cite{FW} and relativistic methods of the "step-by-step" type differ.

Comparing terms in expressions (\ref{eq14erk}) and (\ref{eqgen}),
we can easily establish that they do not completely
agree. Complete agreement occurs for only two terms, determining the series expansion of the operator $\beta\epsilon$ and first of the anticommutators in Eq.
(\ref{eq14gen}), which in turn is the result of transforming the second double
commutator in (\ref{eq28}). These two terms are respectively of the zeroth and first orders in $\hbar/S_0$.
Subsequent terms
in (\ref{eq14erk}) and (\ref{eqgen})
do not coincide. There is a disagreement even for operators proportional to $[{\cal O}^2,[{\cal O}^2,{\cal E}]]$. It is very important that the operators corresponding to them in formulas (\ref{eq14gen})
and (\ref{eqgen}) already appear in
the result of the \emph{first} transformation step, and they must be taken into account in the framework of the
weak-field approximation (${\cal E}\ll m $) when considering terms of the order $(\hbar/S_0)^2$. They arise as a result of
transforming the first double commutator in Eq. (\ref{eq28}). The term $A_{22}$ in (\ref{eq14erk}),
also of the order $(\hbar/S_0)^2$, can
be transformed as
$$\begin{array}{c}
\frac{1}{16m^3}\beta\left\{{\cal O},\left[[{\cal O},{\cal
E}],{\cal E}\right]\right\}=\frac{1}{16m^3}\beta\left[[{\cal
O}^2,{\cal E}],{\cal E}\right]\\-\frac{1}{8m^3}\beta\left([{\cal
O},{\cal E}]\right)^2.
\end{array}$$ Its significant difference from the corresponding term in (\ref{eqgen}) is obvious, as is the difference between the
expressions for the operators $A_{24}$ in the two equations.

In summary, using the relativistic methods of the "step-by-step" type in \cite{JMP,B,PRA,Gos,Gos2,GosHM,GosM},
allows determining
the correct relativistic form of the terms of the zeroth and first orders in $\hbar/S_0$, but the terms of the order
$(\hbar/S_0)^2$ differ from the corresponding terms in the FW representation. Their difference determines the
degree to which the resulting transformation operator obtained by methods of the "step-by-step" type and
equal to the product of the operators of the successive transformations differs from exact FWtransformation
operator (\ref{eq3EI}). The Hamilton operators obtained by the Eriksen method and the methods of the "step-by-step"
type do not agree even for terms of the order
$(\hbar/S_0)^2$. Some terms of the order $(\hbar/S_0)^2$ arise after
the first transformation step and are proportional to the first power of the operator ${\cal
E}$, i.e., they correspond
to the weak-field approximation. Even for them, the Hamilton operators obtained by the Eriksen method
and the methods of the "step-by-step" type do not agree. Such agreement exists if and only if the terms of
second and higher orders in $\hbar/S_0$ under consideration arise as a result of computing the operator $\beta\epsilon$ or the
first anticommutator in (\ref{eq14gen}). For example, the Eriksen method and the iterative methods lead to mutually
consistent results in computing the Darwin interaction, which is of the order $(\hbar/S_0)^2$. This interaction is
determined by the term proportional to
$\Delta\Phi$ in (\ref{eq14ele})and arises as a result of computing the above mentioned
anticommutator.

From results of this study, we can conclude that the iterative methods agree excellently with each
other. Even the nonrelativistic method proposed by Foldy and Wouthuysen
\cite{FW}, with only basic relativistic
corrections taken into account gives expression (\ref{eqoFW}), which is also obtained by the relativistic methods of the
"step-by-step" type (such an agreement does not occur for terms of higher orders in the inverse mass). The
difference between the results obtained from the iterative methods and the results of the Eriksen method,
which realizes the direct FW transformation, is substantially stronger.

According to the analysis of initial Hamilton operator (\ref{2eq3}), the relativistic FW transformation methods
give the form of the transformed Hamiltonian
\begin{equation}\begin{array}{c}
{\cal H}_{FW}=\beta\epsilon+ {\cal E}-\frac 18\left\{\frac{1}
{\epsilon(\epsilon+m)},[{\cal O},[{\cal O},{\cal
F}]]\right\}. \end{array} \label{eqfin}
\end{equation}
This Hamiltonian contains exactly determined terms of the zeroth and first orders in $\hbar/S_0$. Terms of the
second and higher orders in $\hbar/S_0$, if they do not arise as a result of computing Hamiltonian (\ref{eqfin}), cannot
be determined using methods of the "step-by-step" type.

\section{Example: Relativistic particles in a uniform field}\label{example}

As an example showing the importance of methods for the FW transformation of the "step-by-step"
type for relativistic particles, we determine the energy spectrum of spin-0, -1/2, and -1 particles moving
in a plane orthogonal to a uniform magnetic field. For spin-0 particles, the Hamilton operator in the FW
representation has the form \cite{case}:
\begin{equation}
{\cal H}_{FW}=\beta\sqrt{m^2+\bm\pi^{\,2}}, \label{Jeqs}
\end{equation}
and for spin-1/2 particles and the anomalous magnetic moment (AMM)
\cite{JMP,Ts,JETP}, this operator has the form
\begin{equation}
{\cal H}_{FW}=\beta\sqrt{m^2+\bm\pi^{\,2}-e\hbar\bm {\Sigma}
\cdot\bm B}-\mu'\bm\Pi\cdot \bm B. \label{epsilon}
\end{equation} Equations (\ref{Jeqs}) and (\ref{epsilon}) are exact.

If a magnetic field is directed along the $z$, axis and the particle motion is transverse (the eigenvalues of
the operators $p_z$ and $\pi_z$ are zero), then $p_z\!=\!-i\hbar(\partial /\partial z)$
commutes with the Hamiltonian and has eigenvalues
${\cal P}_z\!=\!{\rm const}$. Consequently, considering the particular case ${\cal P}_z\!=\!0$ is well justified
\cite{JMP,JETP}. At the same
time, the problem of particle motion with an arbitrary nonzero eigenvalue ${\cal P}_z$ reduces to this case by a
coordinate transformation.

The energy spectrum of scalar particles is defined by the formula
\begin{equation}
\begin{array}{c} E=\sqrt{m^2+(2n+1)|e|\hbar B},~~~ n=0,1,2,\dots.
\end{array}\label{eq16} \end{equation} We note that for relativistic particles, the two terms in the radicand have the same order of magnitude,
which as a rule requires the condition $n\gg1$.

The energy spectrum and eigenfunctions of AMM particles in constant, uniform magnetic fields were
first found in the Dirac representation \cite{TBZ}. This problem was also successfully solved in the FW representation \cite{DFW}. For ${\cal P}_z\!=\!0$, the formula for the energy spectrum has the form
\begin{equation}      \begin{array}{c}
E=\sqrt{m^2+(2n+1)|e|\hbar B-\lambda e\hbar B}-\lambda \mu'B,
\\ n=0,1,2,\dots,~~~ \lambda=\pm 1.
\end{array}\label{eq16h} \end{equation}

Relations (\ref{Jeqs})--(\ref{eq16h}) show the effectiveness of the relativistic FW transformation by methods of the "step-by-step"
type \cite{JMP} (it is exact in this case). In contrast, the Eriksen method and other direct FW transformation
methods allow representing equations for the Hamiltonian and energy spectrum only by expanding square
roots appearing in them in powers of the operator $|\bm\pi|/m$.

Under the same conditions, we now find the energy spectrum of spin-1 particles whose magnetic
moment has not only the normal part $\mu_0=e\hbar/m$ corresponding to $g=2$ but also the anomalous part
$\mu'=e\hbar(g-2)/(2m)$. The initial Hamilton operator in the Sakata-Taketani representation \cite{SaTa}, derived
in \cite{YB} (see also \cite{EPJC}), in the case under consideration is most conveniently represented as \cite{arXiv}:
\begin{equation}  \begin{array}{c} {\cal H}=\rho_3 {\cal M}+{\cal E}+{\cal
O}, ~~~\rho_3 {\cal E}={\cal E}\rho_3,~~~\rho_3 {\cal O}=-{\cal
O}\rho_3, \\{\cal M}=m+\frac{\bm\pi^2}{2m}-\frac{e\hbar}{m}\bm
S\cdot\bm B, ~~~
{\cal E}=-\rho_3\frac{e\hbar(g-2)}{2m}\bm S\cdot\bm B, \\
{\cal O}=i\rho_2\left[\frac{\bm\pi^2}{2m}-\frac{(\bm\pi\cdot \bm
S)^2}{m}+\frac{e\hbar(g-2)}{2m}\bm S\cdot\bm B\right].
\end{array} \label{eq16u} \end{equation}
Here, the Hamilton operator acts on six-component wave functions $\Psi =\left(\begin{array}{c} \phi \\ \chi
\end{array}\right)$, which are analogues of
bispinors,
$\phi$ and $\chi$ are three-component analogues of spinors, $\bm S$ is a $3\times3$ spin matrix for spin-1 particles, $\rho_i$ are Pauli matrices, and $\rho_i\bm S$ denotes the direct product of matrices, for example,
$\rho_1\bm S\equiv\left(\begin{array}{cc} 0 &
\bm S \\ \bm S & 0 \end{array}\right)$. In the case
under consideration,
$[{\cal M},{\cal E}]=[{\cal M},{\cal O}]=0$, and we can use formulas (\ref{eq14gen}) and (\ref{eq12erk}) replacing $m$ with ${\cal
M}$. In these formulas, the Dirac matrix $\beta$ corresponds to the direct product of $\rho_3$ and the $3\times3$ identity matrix.

For particles without an AMM, the FW transformation is exact, and the obtained Hamiltonian is given
by \cite{EPJC}:
\begin{equation} {\cal H}_{FW}=\rho_3\sqrt{m^2+\bm
\pi^{2}-2e\hbar\bm{S}\cdot\bm B}.  \label{eq1} \end{equation} The energy spectrum has the form
\begin{equation}      \begin{array}{c}
E=\sqrt{m^2+(2n+1)|e|\hbar B-2\lambda e\hbar B}, ~~~
n=0,1,2,\dots,~~~ \lambda=-1,0,+1.
\end{array}\label{eq16e} \end{equation}

For particles with an AMM, the corresponding Hamiltonian was derived in
\cite{JETP2,EPJC} up to terms linear
in the field. To find the spectrum for ${\cal P}_z\!=\!0$, it is convenient to write the Hamiltonian in the form
\begin{equation} {\cal H}_{FW}=\rho_3\sqrt{m^2+\bm
\pi^{2}-2e\hbar\bm{S}\cdot\bm B}-\rho_3\frac{e\hbar(g-2)}{2m} \bm
S\cdot\bm B.  \label{eq1u} \end{equation} With the indicated precision, the energy spectrum is defined by
\begin{equation}      \begin{array}{c}
E=\sqrt{m^2+(2n+1)|e|\hbar B-2\lambda e\hbar B}-\frac{\lambda
e\hbar(g-2)}{2m}B, \\  n=0,1,2,\dots,~~~ \lambda=-1,0,+1.
\end{array}\label{eq16n} \end{equation}

In the case under consideration, the relativistic FW transformation can be accomplished with high
precision, although it is an approximation. The operators in (\ref{eq16u}) satisfy the relations
\begin{equation}      \begin{array}{c}
[{\cal O}^2,{\cal E}]=0,~~~[{\cal O},{\cal E}]=\rho_1\frac{
e^2\hbar^2(g-1)(g-2)}{2m^2}(\bm S\cdot\bm B)^2, \\ \left\{{\cal
O},[[{\cal O},{\cal E}],{\cal E}]\right\}=-\frac{
e^4\hbar^4(g-1)^2(g-2)^2}{2m^4}B^2(\bm S\cdot\bm B)^2, \\
\left([{\cal O},{\cal E}]\right)^2=\frac{
e^4\hbar^4(g-1)^2(g-2)^2}{4m^4}B^2(\bm S\cdot\bm B)^2.
\end{array}\label{eqrel} \end{equation}
In (\ref{eq16u}), among the terms with a nominal order $(\hbar/S_0)^2$, only two, which are proportional to
$\left\{{\cal O},[[{\cal O},{\cal E}],{\cal E}]\right\}$ and
$\left([{\cal O},{\cal E}]\right)^2$, are nonzero, and they are in fact of the fourth order in $\hbar$ and $B$.
Because $S_0=\epsilon^2/(|e|B)\sim
m^2/(|e|B)$ in the considered case, the relativistic FW transformation allows determining the Hamiltonian for
spin-1 particles up to terms of the order $(|e|\hbar B)^3/m^5$ inclusively. With these terms taken into account, the
Hamiltonian becomes
\begin{equation}\begin{array}{c} {\cal H}_{FW}=
\rho_3\epsilon-\rho_3\frac{e\hbar(g-2)}{2m} \bm
S\cdot\bm B\\+\rho_3\frac{e^2\hbar^2(g-1)(g-2)}{16m^3}
\Biggl\{\frac{1}{\epsilon(\epsilon+m)},\biggl(B^2(\bm S\cdot\bm
\pi)^2\\-[\bm S\cdot(\bm \pi\times\bm B)]^2-e\hbar(g-1)B^2(\bm
S\cdot\bm B)\biggr)\Biggr\},\\\epsilon=\sqrt{m^2+\bm
\pi^{2}-2e\hbar\bm{S}\cdot\bm B-\frac{e^2\hbar^2g(g-2)}{4m^2}(\bm
S\cdot\bm B)^2}.
\end{array}\label{eqprf}
\end{equation}

We note that the effects due to the tensor electric and magnetic polarizability indicated in the introduction
are determined by spin-dependent terms of the second order in
$\hbar$ and $B$. For these terms, we have
the equality
$$B^2(\bm S\cdot\bm
\pi)^2+[\bm S\cdot(\bm \pi\times\bm B)]^2+\bm \pi^2(\bm S\cdot\bm
B)^2=2(\bm \pi\times\bm B)^2.$$ The term in the right-hand side characterizes the scalar electric polarizability of a moving particle.

The definition of the Hamiltonian and energy spectrum for spin-1 particles with an AMM in a homogenous
magnetic field is also a demonstration of the possibilities provided by the iterative methods
for the FW transformation. Neither in the Sakata–Taketani representation nor with other transformation
methods (see \cite{Ts,TY}) this problem has been solved. Moreover, the problem of consistency of the quantum
mechanics of spin particles had been discussed for several years (see \cite{Ts,GoldmanTsaiY,VSM} and the references
therein).

\section{Discussion and conclusions}

Comparing the results obtained by the Eriksen method and by methods of the "step-by-step" type
leads to an important and rather surprising conclusion. Relativistic FW transformation methods of the
"step-by-step" type already after the \emph{first} transformation step (let alone the subsequent steps) give an
expression for the Hamiltonian inconsistent with the exact result determined from a series of relativistic
corrections computed by the Eriksen method.

The disagreement between the Eriksen method realizing the direct FW transformation and methods
of the "step-by-step" type occurs for terms of the second and higher orders in $\hbar/S_0$, if they do not arise as
a result of computing the operator
$\beta\epsilon$ or the first anticommutator in
(\ref{eq14gen}). The exact Eriksen method and the
relativistic methods of the "step-by-step" type developed in \cite{JMP,B,PRA,Gos,Gos2,GosHM,GosM}
lead to fully consistent results only for terms
arising as a result of computing these operators, including all terms of the zeroth and first orders in $\hbar/S_0$.

The Eriksen method allows representing the transformed Hamiltonian as an expansion in powers of
the operators ${\cal
E}/m$ and ${\cal O}/m$. Therefore, it is convenient to use it to solve nonrelativistic problems (e.g.,
in atomic physics) when relativistic corrections must be taken into account. For relativistic methods of the
"step-by-step" type, the operators ${\cal E}$ and ${\cal O}$ are not considered small, and the expansion is based on the
assumption that the commutator of these operators is small in order of magnitude compared with their
products. The validity of this assumption depends on the problem considered. It is quite often invalid in
atomic physics but is always valid in the quantum mechanical description of particle beams in accelerators,
storage rings, and Penning traps. On the other hand, the Eriksen method not only does not allow deriving
compact relativistic expressions for the Hamiltonian in the FW representation but plainly gives a divergent
series of relativistic corrections when $\bm
p^2/m^2\ge1$. Therefore, there is currently no alternative to iterative
methods for describing relativistic particles.

In particular, with adequate precision, iterative methods describe spin effects for relativistic particles
including spin affecting the motion trajectory (determined by the respective Stern-Gerlach and Mathisson
forces for electromagnetic and gravitational interactions \cite{JMP,PRA,OST,OSTRONG}). These effects are determined by
terms of the order $\hbar/S_0$. There are some indirect arguments for extending this conclusion to spin-($s\neq1/2$) particles. Terms linear in the spin and Planck constant in the FW representation Hamiltonian derived in \cite{JETP2} (see also \cite{EPJC}) by a method of the "step-by-step" type for spin-1 particles lead to an operator equation for
the spin motion that corresponds to the Thomas-Bargmann-Michele-Telegdi equation and consequently
correctly describes spin effects. The effectiveness of iterative methods was shown in Sec. 5 in the example
of relativistic spin-0, -1/2, and -1 particles in a homogenous magnetic field. Using the results obtained here,
we determined the precision provided by methods of the "step-by-step" type in that case.

In the general case, the problem of correctly defining the contributions to the Hamiltonian determined
by the quadrupole interaction and tensor magnetic and electric polarizability of spin-($s>1/2$) particles and
having the order $(\hbar/S_0)^2$ by a method of the "step-by-step" type needs separate study. Additional analysis is
needed because the initial Proca-Corben-Schwinger equation and similar equations for higher-spin particles
characterize structureless point particles.

A comparison of the three methods of the "step-by-step" type
\cite{JMP,B,PRA,Gos,Gos2,GosHM,GosM} with the most fundamental
justification showed the principal commonality of the approaches used. At each step, a transformation
operator is chosen that would realize the exact transformation if the even and odd operators commute.
This commonality leads to agreement of the results of the three methods. But a substantial difference in
quantum mechanics formalisms used in these methods leads to differences in the amount of computation
needed to obtain the final results. The simplest way to derive the Hamiltonian is to use the traditional
operator formalism and unitary transformations \cite{JMP,PRA} computing in accordance with (\ref{eqfin})
and estimating
the computational precision (determining the order of magnitude) by computing the neglected terms in
accordance with
(\ref{eq14erk}). An example of this approach was given in Sec.
5, and this example also showed the
possibility of using the method for particles with spins other than 1/2.

The precision estimation allows detecting the difference of iterative methods from Eriksen's formal
expansion of the Hamiltonian and shows their applicability boundaries. We note that computing by (\ref{eqfin})
requires much fewer computational steps than using the other two iterative methods analyzed here. Nevertheless,
the method developed and applied in
\cite{Gos,Gos2,GosHM,GosM,Gosgrap} is the best for studying effects determined by Berry
phases. We also note a successful use of a similar method in \cite{BliokhK}, also for the analysis of Berry phases.
\medskip

\textbf{Acknowledgments.} The author is deeply grateful to Professor K. A. Sveshnikov for his remarks.
This work was supported by the Belorussian Republican Foundation for Basic Research (Grant No.
$\Phi$12D-002).

\end{document}